\documentstyle[10pt,aaspp4,psfig]{article}

\newcommand\etal {\hbox{et al. }}

\righthead{Time Delays in kHz QPO}

\begin{document}

\title{Discovery of microsecond time lags in kilohertz QPO}

\author{B. A. Vaughan} 
\affil{Space Radiation Laboratory, California Institute
of Technology, MC 220-47, Pasadena CA 91125; brian@srl.caltech.edu}

\author{M. van der Klis, M. M\'endez\altaffilmark{1}, 
J. van Paradijs\altaffilmark{2} and R. A. D. Wijnands}
\affil{Astronomical
Institute ``Anton Pannekoek'', University of Amsterdam, \\ and Center
for High Energy Astrophysics, Kruislaan 403, NL-1098 SJ Amsterdam, The
Netherlands; michiel@astro.uva.nl, mariano@astro.uva.nl, jvp@astro.uva.nl,
rudy@astro.uva.nl}

\author{W.H.G. Lewin} \affil{Massachusetts Institute of Technology,
Cambridge, MA 02139; lewin@space.mit.edu}

\author{F.K. Lamb and D. Psaltis} \affil{Departments of Physics and Astronomy,
University of Illinois at Urbana-Champaign, Urbana, IL 61801;
f-lamb@uiuc.edu, demetris@astro.uiuc.edu}

\author{E. Kuulkers} \affil{Astrophysics, University of Oxford, Nuclear and Astrophysics
Laboratory, Keble Road, Oxford OX1 3RH, United Kingdom;
e.kuulkers1@physics.oxford.ac.uk}

\author{T. Oosterbroek} \affil{Astrophysics Division, Space Science Department of ESA, 
ESTEC, P.O. Box 299, 2200 AG Noordwijk, \\
The Netherlands; toosterb@estsa2.estec.esa.nl}

\altaffiltext{1}{Also Facultad de Ciencias Astron\'omicas y Geof\'{\i}sicas, 
   Universidad Nacional de La Plata, Paseo del Bosque S/N, 
   1900 La Plata, Argentina}

\altaffiltext{2}{Also Department of Physics,
University of Alabama at Huntsville, Huntsville, AL 35899}

\begin{abstract}

Using the Rossi X-ray Timing Explorer 
we have measured $\sim$27$\mu$s time delays in 830\,Hz 
quasi-periodic oscillations (QPO)
between 4--6\,keV and 11--17\,keV in 4U~1608--52, 
with high-energy photons lagging low energy photons, and 
found upper limits to
the time delays of 45$\mu$s between 2--6.5\,keV and 6.5--67\,keV in 
$\sim$730\,Hz QPO
in 4U~0614+091 and 30$\mu$s between 8.7--12.4\,keV and 12.4--67\,keV in 
$\sim$870\,Hz QPO in 
4U~1636--53.  We also find that the cross coherence function
between QPO at different energies is $>0.85$ with
95\% confidence in 4U~1608--52 and 4U~1636--53.
If Compton upscattering of low-energy X-rays in a region with an optical
depth of a few is responsible for the delays,
then the Compton upscattering region is between a few kilometers and
a few tens of kilometers in size.

\end{abstract}


\section{Introduction \label{intro}}

Quasi-periodic oscillations (QPO) with frequencies 
$\nu_{\rm QPO}\sim350-1170$\,Hz have recently
been observed in power spectra of 
countrate modulations in $\sim$10 X-ray binaries.
These QPO have fractional root-mean square (rms) amplitudes of 1--20\%, and 
quality factors $Q\equiv\nu_{QPO}/\Delta\nu_{\rm QPO}=10-200$,
where $\Delta\nu_{\rm QPO}$ is the full width at half maximum of the QPO
peak in the power spectrum.  In
six sources a pair of QPO peaks have been observed simultaneously
(4U~0614+091, 4U~1728--34, Sco~X-1, GX~5--1, 4U~1820--30, and 4U~1636--53), 
with frequencies 
separated by 200--400\,Hz.  In 4U~0614+091 and 4U~1728--34
the frequency separation remains constant during excursions in
QPO frequency by $\sim$200\,Hz.  QPO frequency is strongly correlated
with count rate in 4U~1820--30, 4U~0614+091, and
4U~1728--34, and with $\dot M$ as inferred from the Z-track 
in GX~5--1 and Sco~X-1 (van der Klis \etal 1996).
  A 363\,Hz oscillation seen during 
X-ray bursts in 4U~1728--34 (Strohmayer \etal 1996)
has a frequency consistent with the difference
between the pair of QPO peaks.

The frequencies of these QPO may correspond to the
Keplerian frequency, $\nu_{\rm Kep}$, at the inner edge of the
Keplerian flow, which may be terminated by radiation forces,
general relativistic corrections to Newtonian gravity, or
perhaps the magnetic field of the neutron star and to the
difference, or beat, frequency between $\nu_{\rm Kep}$ and the
spin frequency of the neutron star (Alpar \& Shaham 1985; Lamb et
al. 1985; Miller, Lamb, \& Psaltis 1997). The oscillations seen
in \hbox{4U~1728$-$34} during some bursts, which have a frequency
equal to the difference between the frequencies of the two
higher-frequency QPO, suggests this interpretation. The changing
difference between the frequencies of the high-frequency QPO
seen in \hbox{Sco~X-1} (van der Klis et al.\ 1996) would then
require the spin frequency to beat with a varying Keplerian
frequency that is slightly different from the frequency of the
higher-frequency QPO in this source.

In the Z sources GX~5--1 and Cyg~X-2, 
countrate modulations at energies above $\sim$5\,keV lag those
below $\sim$5\,keV by 1--10\,ms in 15--55\,Hz horizontal-branch QPO
(van der Klis \etal 1987; Vaughan \etal 1994), 
with the delay
increasing with photon energy,
and by 70\,ms in
6\,Hz normal-branch QPO (Mitsuda \& Dotani 1989; Vaughan \etal 1997;
Dieters \etal 1997).  
In both cases, the cross coherence between oscillations at different
energies is $\sim$1, meaning that they can be related to one another
by a constant linear transformation.  See Vaughan \& Nowak 1997 for
an explanation of the cross coherence function, also called the coherence
function, or simply coherence.
No lags have so far been detected 
in normal-branch QPO in Sco X-1 (Dieters \etal 1997).
In the case of 
horizontal-branch oscillations, Compton upscattering of low-energy photons has
been suggested as one possible explanation for the lags 
(Wijers, van Paradijs \& Lewin 1987; Stollman \etal 1987; Bussard \etal 1988), 
although differences
in the lag at the QPO fundamental frequency and its harmonic in GX~5--1
challenge this interpretation (Vaughan \etal 1994).  
It is also possible to interpret the
lags in the context of shot noise models by postulating that individual
shots ``harden'', i.e., become hotter, as they progress.
  Hardening shots
can be used to produce any time delay, but
Vaughan \& Nowak
(1997) point out that for blackbody emission,
hardening shots only result in unity cross coherence
if the photons are emitted in the Rayleigh-Jeans part of the energy spectrum,
in which case they lead to no delay.  
We report in this paper the discovery of $\sim 27$~$\mu$s hard
lags in kilohertz QPO and show that they are consistent with
Compton upscattering in a region between a few kilometers and few tens
of kilometers in size, i.e., comparable
to the size of a neutron star.

\section{Observations  \label{observations}}

We have investigated time delays in the QPO between different energy channels
in 4U~1608--52, 4U~0614+091, and 4U~1636--53.  These sources were chosen because
each showed QPO with high fractional rms amplitude, a moderate to high count 
rate, a narrow 
peak, and a stable frequency during at least one extended interval.
We also of course required spectrally-resolved data of sufficient time
resolution.

All observations were performed using the 
Proportional Counter Array on the
Rossi X-ray Timing Explorer (Bradt, Rothschild \& Swank 1993).  
Observation times and durations, and count rates are given in Table 1, 
along with QPO properties
(all of which were measured by previous investigators).
Details of the QPO in these sources can be found in Berger \etal 1996
(4U~1608--52), M\'endez \etal 1997 (4U~0614+091), and Wijnands \etal 1997
(4U~1636--53).
A single QPO peak was present in each observation.  We were unable to investigate
time lags in observations containing two QPO peaks because data with sufficient
signal strength were unavailable (see section \ref{results}).
We calculated the cross-spectra in the way described in Vaughan \etal (1994).

{\begin{table}[tb]
\vspace*{-0.0in}
\caption{Observations and QPO Properties}
\vspace{-0.1in}
\begin{center}
\begin{minipage}{6.5in}
\renewcommand{\thempfootnote}{\alph{mpfootnote}}
\renewcommand{\thefootnote}{\alph{footnote}}
\renewcommand{\footnoterule}{}
\def\fm{\footnotemark}
\begin{tabular}{lcccclc} \hline \hline \\
\multicolumn{1}{c}{System} & 
\multicolumn{1}{c}{Observation Date} & 
Duration (s)\fm[1] & 
Rate (s$^{-1}$)\fm[2]
& $\nu_{\rm QPO}$ (Hz) &
\multicolumn{1}{c}{RMS}\fm[3] &
Range (Hz)\fm[4] \\ \hline
4U 1608--52 &  3 March 1996 & 3694 & 3100 & 830 & .07 & 820--840 \\
4U 0614+091 & 16 March 1996 & 4918 & 560  & 730 & .13 & 680--780 \\
4U 1636--53 & 29 May   1996 & 2360 & 1700 & 870\fm[5] & .07 & 862--877 \\
\hline
\end{tabular}
\vskip -0.1cm
\footnotetext[1]{Length of data set used for analysis.  Determined principally
by the stability of the QPO frequency}
\footnotetext[2]{Average count rate of data used in the analysis}
\footnotetext[3]{Fractional RMS amplitude of the QPO.  Increases with energy in all 
sources}
\footnotetext[4]{Range of frequencies used for analysis.  Dictated by the
QPO width and stability of the QPO centroid frequency}
\footnotetext[5]{Average $\nu_{\rm QPO}$ for the first 2360\,s of the observation}
\end{minipage}
\end{center}
\end{table}
}

Instrumental deadtime can cause a bias in measurements of time delays by
introducing an anti-correlation between pairs of energy channels that manifests
itself as a 180$^\circ$ phase difference in the cross spectrum between
pairs of energy channels at all frequencies.  Because the count rates for each
of the sources investigated here were small compared with the inverse of 
the instrumental
deadtime, the fractional deadtime was
typically $<$1\% and deadtime had a completely 
negligible effect on cross spectra measured at the QPO frequency.

\section{Results \label{results}}

Measurement of time delays and the cross coherence function
in the presence of a counting-noise
background is discussed in detail in Vaughan \etal (1994) and Vaughan \& Nowak (1997).  The 
crucial quantity for determining whether 
a meaningful estimate of the time delay
between two light curves is possible is
the signal to noise ratio, $S/N$, given by
$S/N=f_1f_2r_1^{1/2}r_2^{1/2}T^{1/2}\Delta\nu^{-1/2}$,
where $f_1$ and $f_2$ are the fractional rms amplitude of the QPO
in the 2 light curves, $r_1$ and $r_2$ are the count rates, $T$ is the
duration of the measurement, and $\Delta\nu$ is the width of the frequency
interval used for the measurement.  This formula assumes the background
count rate is much smaller than the source count rate and that the cross
coherence between the light curves is unity.  The smallest time difference that
can be measured 
is $\delta t_{\rm min}=
n_\sigma(2\pi\nu)^{-1}[\arctan(S/N)]^{-1}
\approx n_\sigma(2\pi\nu)^{-1}(S/N)^{-1}$, where $n_\sigma$
is the detection significance, in standard deviations.

\centerline{\hbox{\psfig{file=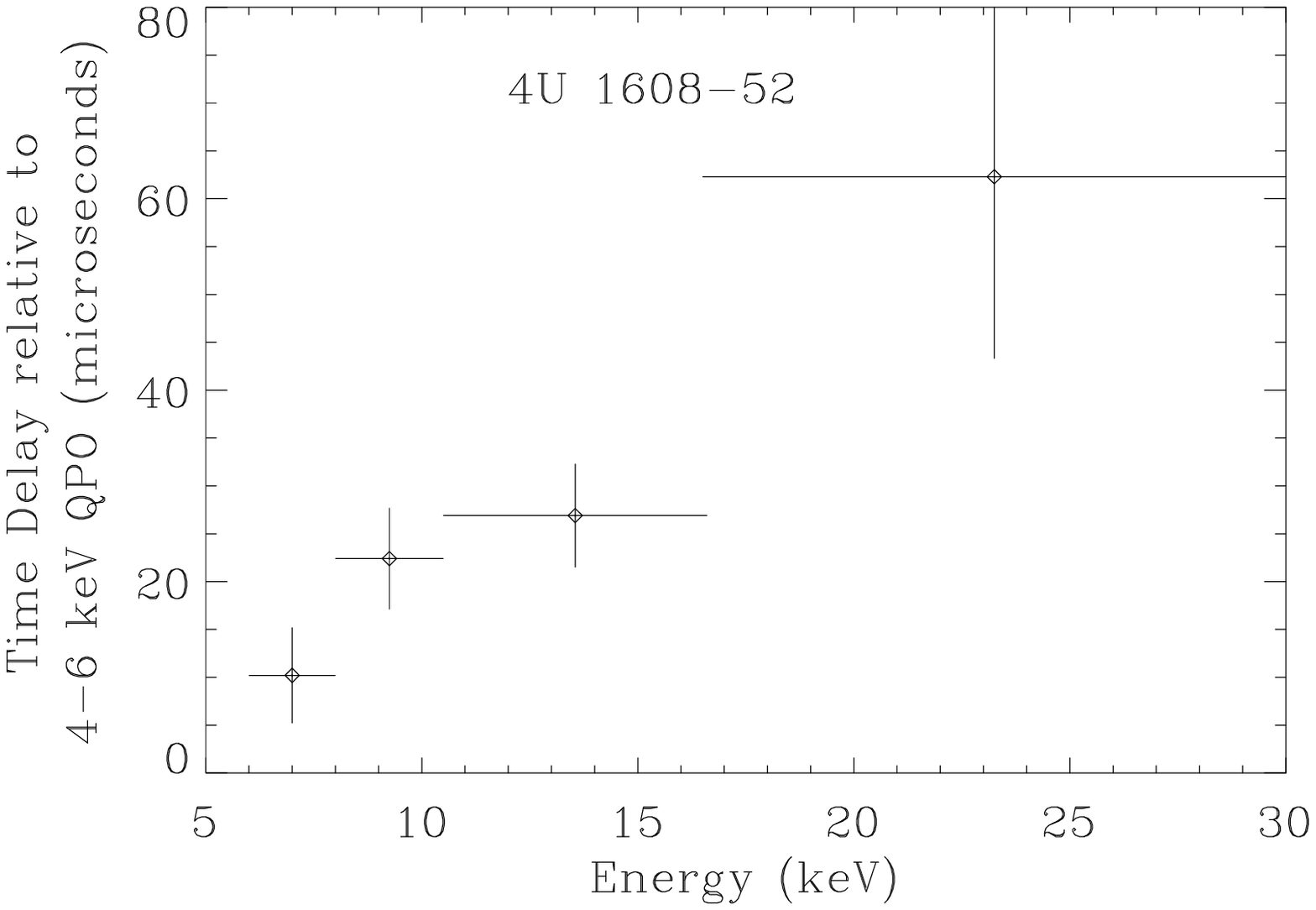,height=3in}}}
\figcaption{
Time delays in 830\,Hz QPO in 4U~1608--52 relative to 4--6\,keV fluctuations, 
measured in the frequency range 820--840\,Hz on 1996 March 3.}

The observation from Table 1
best suited for measuring time delays is the 3 March 1996
observation of 4U~1608--52, for which $S/N\sim 40$ between energy channels
of width $\sim$2\,keV at energies below $\sim$12\,keV.  We divided the data
containing strong $\sim830$\,Hz QPO
into 5 energy channels covering the ranges 4--6\,keV, 6--8.2\,keV,
8.2--10.6\,keV, 10.6--16.6\,keV, and 16.6--30\,keV, and measured the
time delay and cross coherence function
between all pairs of channels.  The time
delays are plotted in Figure 1 relative to the 4--6\,keV channel.  
The sign convention is such that a positive time delay indicates that
the higher energy channel lags behind the lower energy channel. The
8.2--10.6, 10.6--16.6, and 16.6--30\,keV channels all lag
the 4--6\,keV channel with $>3\sigma$ significance.  The
most significant lag (5$\sigma$) is between the 4--6\,keV and 10.6--16.6\,keV channel:
$27\pm5\mu$s.  The cross coherence function between the channels
is consistent with unity, and is $>0.9$ with 95\% confidence for all channels.

We are also able to significantly constrain the time delay in
the single QPO peak at $\sim730$\,Hz observed 16 March 1996 in
4U~0614+091, and in the single QPO peak at $\sim870$\,Hz observed
29 May 1996 in 
4U~1636--53, as well as the cross coherence in 4U~1636--53.  In both 
cases, to optimize $S/N$, we divided the data into 2 energy channels.
In \hbox{4U~06214$+$091} we find that the oscillations at
6.5--67~keV lag those at 2--6.5~keV by $26 \pm 23$~$\mu$s and
in \hbox{4U~1636$-$53} we find that the oscillations at
12.4--67~keV lag those at 8.7--12.4~keV by $-2.7 \pm
14$~$\mu$s.
For these two systems we thus have 95\% confidence upper limits
on the time delay, $\delta t$, of 45\,$\mu$s (4U~0614+091) and 30\,$\mu$s
(4U~1636--53).  The 95\% confidence lower limit to the cross coherence
is 0.85 in 4U~1636--53.  The cross coherence function 
of the QPO in 4U~0614--091 could not be
meaningfully constrained since cross 
coherence is a 4th-order statistic with a large variance
at small $S/N$.  We attempted to investigate time lags in 
observations containing two simultaneous 
kilohertz QPO peaks.  Our best candidate was a  
28 February 1996 observation of 4U~1636--53
with QPO at 900\,Hz and 1176\,Hz (Wijnands \etal 1997).
We were unable to measure the time delay in
either of the two QPO peaks due to low $S/N$.

The small lags we see cannot be due to large ($\delta t \gg 1/\nu$) and 
variable time differences 
that average out to zero.  If the lags were large and variable, 
we would find phase differences $\delta\phi=2\pi\nu\delta t$
uniformly and randomly distributed on $[-\pi,\pi]$, large error 
bars $\Delta(\delta\phi)\sim\pi$, and small cross coherence.  

\section{Discussion \label{discussion}}

We have measured time delays between the kilohertz QPO at
different photon energies in 4U~1608$-$52, 4U~0614$+$091,
and 4U~1636$-$53. In the first source we found a significant
hard lag whereas in the latter two sources we found that any
time delays are $\lesssim 45~\mu$s. The QPO investigated
here have fractional rms amplitudes and widths
comparable to those of the horizontal-branch oscillations
in the Z sources but their frequencies are 10--40 times
higher, so we are sensitive to time lags 10--40 times
smaller. In 4U~1608$-$52, the $\sim 27~\mu$s time lag is equivalent
to a phase difference of $\sim 0.13$~radians, or
about $8^{\circ}$.

The time lags we have measured can be used to derive general
constraints on the X-ray production mechanisms in these
systems (see Lamb 1988). 
It is important to consider these constraints in
the larger context of the time lags, energy spectra, power spectra, and
cross coherence between the oscillations at different photon
energies. Any physical model should be consistent with all
four of these statistical properties. If the X-ray spectrum
is not formed predominantly by Compton scattering, then the
$\sim 800$~Hz oscillation at 11--16~keV must be produced so
that it lags the oscillation at 4--6~keV by $\sim 27~\mu$s
in 4U~1608$-$52.  Delays must be $\lesssim 45~\mu$s in 4U~0614$+$091
and 4U~1636$-$53. The high value ($\gtrsim 0.85$) of the cross
coherence in 4U~1636$-$53 and 4U~1608$-$52 requires that if the 
source of photons responsible for the QPO is physically extended, 
the lags in the emission process must be independent of location
(Vaughan \& Nowak 1997; Nowak \& Vaughan 1996).  Any model in 
which the QPO are produced at the surface of a neutron
star that is also a pulsar must explain why the QPO are seen 
but the periodic oscillations are hidden. 

We now consider models in which photons produced near the
neutron star are upscattered by electrons in a
region with optical depth $\tau \gtrsim 1$. 
Optically thin free-free emission in a scattering region
$\lesssim$100 km in size supplies too few
photons to account for the observed luminosity
of the atoll sources, which is  $\sim$0.1 $L_{\rm Edd}$.
Hence, the photons must be produced in a
self-absorbed region, such as the outer layers of the neutron
star. Given the observed luminosities of these sources, any
thermal emission from the neutron star peaks at $\sim$1--2 keV. No
such peak is observed, indicating that the scattering region has
an optical depth of $\gtrsim 3$.  Models of this type
provide a natural explanation for the X-ray spectra of
these sources (see Lamb 1989; Psaltis, Lamb, \& Miller 1995)
and may also offer an explanation
of their rapid x-ray variability (see Miller, Lamb, \& Psaltis 1997 for an
overview).  We analyze the generic properties
of models of this type, avoiding special geometries or
physical conditions. 

The fractional rms amplitudes of the kilohertz QPO can be used to 
place an upper bound on the sizes of emitting
regions in all kilohertz QPO sources. 
Quite generally, the size of the
region must be less than a few times $c/\nu_{\rm QPO}$,
independent of the QPO mechanism, since radiation
from different parts of the region have different path lengths to the 
observer, causing phase shifts that attenuate the QPO signal. 
If the radiation is scattered, the rms amplitude at
infinity of a luminosity oscillation with frequency $\nu$
and amplitude $A_0$ at the center of a spherical 
region of radius $a$ and optical depth $\tau$ is $A_\infty
\simeq (2^{3/2}x e^{-x}+e^{-\tau})A_0$, where $x\equiv (3\pi
\nu a\tau/c)^{1/2}$ (Kylafis \& Phinney 1989). The $\sim
800$~Hz QPO studied here have rms amplitudes $\sim 10\%$,
so if they are luminosity oscillations the scattering
region must be
smaller than $\sim 200$~km if $\tau\sim5$. If instead they are
beaming oscillations, the scattering region must be smaller
than $\sim
100$~km if $\tau\sim5$. These bounds are almost independent
of the electron temperature and bulk velocity in the
scattering region, and remain approximately valid for non-spherical
regions.  

The time lags reported here, if caused by upscattering,
provide stringent constraints on the 
size, $a$, of the upscattering region.  We first compute a
rough, qualitative estimate of $a$.
If photons with energy $E_{\rm in}$ are injected into a
scattering region where
the electron temperature $T_{\rm e}$ is $\gg E_{\rm in}/k_B$, 
they will gain energy
each time they scatter, so photons that escape
with a higher energy $E_2$ emerge later than photons
that escape with a lower energy $E_1$. The delay $\delta t$
in the arrival time at a distant observer
of photons of energy $E_2$ relative to
photons of energy $E_1$ depends on the geometry of the
region and the spatial distribution of its electron density
and temperature but will be $\sim \Delta u\, \ell/c$, where
$\ell$ is the photon mean free path and $\Delta u$ is
the difference in the average number of scatterings
experienced by photons that emerge with energies $E_1$ and
$E_2$. If $\Delta u \sim u$, where $u$ is the average number of 
scatterings to escape from the cloud, then $\delta t\sim
a u/c\tau$, where $\tau = a/\ell$ is the optical depth.
For $\delta t \sim
27~\mu$s, $\tau \sim 5$, and $u\sim\tau^2$,
the inferred size of the
scattering region is a few kilometers.

To estimate the size of the region more quantitatively, note
that if the scattering electrons are nonrelativistic, then
after one scattering the average energy of a photon with
initial energy $E_{\rm in}$ is
$ E\sim E_{\rm in} \exp(4 kT_e/m_e c^2)$, 
 where $m_e$ is the electron rest mass (see Rybicki \&
Lightman 1979, Ch.~7). Therefore, the ratio of the energies
of two photons that experience a different number of
scatterings is $E_2/E_1 \sim \exp(4\Delta u k T_e/m_e
c^2)$. Solving this expression for $\Delta u$ and using the
above expression for $\delta t$ gives (see, e.g., Sunyaev
\& Titarchuk 1980)
 \begin{equation}
 \delta t \sim
 \frac{a}{c\tau}\frac{m_e c^2}{4 k T_e}
 \ln\left(\frac{E_2}{E_1}\right)\;.
 \end{equation}
 For 4U~1608$-$52, we have measured $c\,\delta t$ to be $\sim 8$~km for
$E_1=5$~keV and $E_2=14$~keV, and hence $a \sim
8\tau(4kT_e/m_e c^2)~{\rm km}$.

Photons injected with energies $E_{\rm in} \ll kT_e$ into a
region with a Compton parameter $y \equiv 4
kT_e \tau^2/(m_ec^2)$ that is less than the saturation value
$y_{\rm sat} \sim 10$ emerge with a spectrum that is
roughly a power law with an exponential cutoff at $\sim 2
kT_e$. The countrate spectra of the atoll sources indicate
$kT_e \gtrsim 10$~keV (White, Stella, \& Parmar 1988), in
which case $a$ must be greater than $\sim \tau \cdot 1$~km.
Compton upscattering produces hard time lags only if it is
unsaturated, i.e., only if $y \lesssim y_{\rm sat}$, so $a$
must also be less than $\sim 8\, (y_{\rm sat}/\tau)$~km.
Combining these upper and lower limits, we find
 \begin{equation}
 5 \left(\frac{\tau}{5}\right) \;\mbox{km}
 \lesssim a \lesssim
 16 \left(\frac{y_{\rm sat}}{10}\right)
 \left(\frac{5}{\tau}\right) \;\mbox{km} \;,
 \end{equation}
 where we have scaled the expressions to $\tau \sim 5$, the
value suggested by models of the X-ray spectra and rapid
X-ray variability of the atoll sources (see for example
Lamb 1989; Psaltis, Lamb, \& Miller 1995).

In this analysis we have assumed that soft photons are
injected at the center of a spherical region of uniform
temperature and electron density. If any of these
conditions are not satisfied, the bounds on the size of the
Compton upscattering region are larger than those derived above. We
have also assumed that $E_2 \ll 2kT_e$. If instead $E_2
\sim 2kT_e$, the energy change in each scattering of photons
with energy near $E_2$ is significantly less than we have
assumed, so $\delta t$ is significantly greater than in
equation~(1) and the bounds on $a$ are correspondingly
smaller than estimated above. However, the value of $E_2$
used here (14~keV) is less than the $\sim 20$~keV lower
bound on $2kT_e$ inferred from the X-ray spectra of the
atoll sources, so this effect is probably small.
We expect that the upscattering region is in reality
inhomogeneous and that the size of the upscattering region is
closer to the lower bound rather than the upper bound in
inequality~(2) (see Miller et al. 1997). This is natural since we
know the 4U~1608--52 is a neutron star because it produces
Type~I X-ray bursts (see Brandt et al. 1992 and references therein).

Compton upscattering by electron orbital motions may also be
important in forming the X-ray spectra of these sources. If
the radiation field is nearly isotropic and the typical orbital
velocity of the electrons is $V$, the spectrum that emerges
from the scattering region is roughly a power law up to a
cutoff energy $E_c \sim m_e c^2(V/c)^2$ (Psaltis \& Lamb 1997).
For example, if $V/c\sim 0.15$, which can easily be attained in 
regions as close to
the neutron star as we are considering, then $E_{\rm c}\sim10$\,keV.
If Compton upscattering by electron bulk motion
is important, the time lags will depend upon
the radial dependence of $V$, but the bounds on the size of
the scattering region remain about the same. 

In summary, the time lags we have measured indicate that
the Compton upscattering region in \hbox{4U~1608$-$52} is 
between a few kilometers and a few tens of kilometers in
size, i.e., comparable to the size of a neutron star.

\acknowledgments

The authors acknowledge Michael Nowak and Lev Titarchuk
for useful comments and suggestions.
BAV acknowledges support from the United States National Aeronautics and Space 
Administration (NASA) under grants NAG 5--3340 and NAG 5--3293.
JvP acknowledges support from
NASA under grant NAG 5--3271. FKL and DP acknowledge NSF grants AST 93--15133 
and AST 96--18524 and NASA grant NAG 5--2925.
MM is a fellow of the Consejo Nacional de Investigaciones
Cient\'{\i}ficas y T\'ecnicas de la Rep\'ublica Argentina.
This work was supported in part by the Netherlands Organization for
Scientific Research (NWO) under grant PGS 78--277 and by the Netherlands 
Foundation for Research in Astronomy (ASTRON) under grant 781--76--017.

\end{document}